\documentclass[aps,prb,twocolumn,showpacs,groupedaddress]{revtex4}

\usepackage{amsmath}
\usepackage{graphicx}
\usepackage{dcolumn}
\usepackage{bm}

\begin{document}

\title{Superfluidity vs. localization in bulk $^{\bf 4}$He at zero
temperature}

\author{C. Cazorla}
\author{J. Boronat}
\affiliation{Departament de F\'\i sica i Enginyeria Nuclear, Campus Nord
B4-B5, Universitat Polit\`ecnica de Catalunya, E-08034 Barcelona, Spain}

\date{\today}
\begin{abstract}
We present a zero-temperature quantum Monte Carlo
calculation of liquid $^4$He immersed in an array of confining potentials.
These external potentials are centered in the lattice sites of a fcc solid
geometry and, by modifying their well depth and range, the system evolves
from a liquid phase towards a progressively localized system which mimics a
solid phase. The superfluid density decreases with increasing order,
reaching a value  $ \rho_{\rm s}/\rho = 0.079(16)$ when the Lindemann's ratio of the
model equals the experimental value for solid $^4$He.
\end{abstract}


\maketitle

\section{Introduction}

The counterintuitive coexistence of superfluidity and solid order has been
the object of theoretical and experimental debate for a long time.  The
recent experimental findings by Kim and Chan of non-classical rotational 
inertia (NCRI) in solid $^4$He, both in bulk~\cite{moses1} and 
in a confined environment,\cite{moses2,moses2_bis} suppose  a real  breakthrough in the field. Using a torsional
oscillator setup and ultrahigh-purity $^4$He, a superfluid density
$\rho_s/\rho \sim 1-2$\% has been measured below a temperature $T=230$mK.
The transition temperature from normal solid to supersolid is nearly
constant with the pressure and no signal of the transition is observed in
the specific heat.\cite{moses3}  On the other hand, other manifestations of superfluidity
like pressure-induced flow has not been detected in a recent experiment
where solid $^4$He was confined in Vycor.\cite{day}

On the theoretical side, the pioneering works by Andreev and
Lifshitz~\cite{andreev} and
Chester~\cite{chester} suggested for the first time the possibility of Bose-Einstein
condensation of vacancies or interstitials in solid $^4$He. Recently,
Prokof'ev and Svistunov~\cite{proko} have worked out on the same idea and have
established that the existence of defects in the solid is a necessary
condition for observing superfluidity, i.e., the supersolid phase.
According to this mechanism for superfluidity in the solid phase, $^4$He
could present at very low temperature the form of an incommensurate
crystal, i.e., a quantum solid where the number of lattice sites is
different  from the number of atoms.\cite{anderson} 
Also, the possibility of a disordered
solid phase (glass) has been suggested as an alternative for
observing superfluid signals.\cite{glass}  However, the hypothesis of incommensurability
contradicts well established experimental data  about the absolutely
negligible concentration of vacancies at temperatures where supersolid has been
observed.\cite{meisel}

Following Legett's arguments,\cite{legett}  a second plausible scenario for a supersolid
phase would be the one of a commensurate crystal but with enough exchanges
between atoms to generate NCRI. This fascinating possibility can only be
thought in a quantum solid such as $^4$He where the displacements of the
atoms around the lattice sites are incredibly large from a classical point
of view. However, recent path integral Monte Carlo (PIMC) simulations of
solid $^4$He near the experimental transition temperature have not detected
neither superfluid density~\cite{ceper1} nor off-diagonal long-range
order.\cite{ceper2}  It is worth
mentioning that the estimation of both magnitudes is rather difficult using
PIMC since the temperature is very low and the signal is expected to be very
small.  

Zero-temperature properties of solid $^4$He such as its equation of state,
static structure factor or Lindemann's ratio have been calculated using
diffusion Monte Carlo (DMC)~\cite{leandra} and Green's function Monte
Carlo (GFMC).\cite{kalos}  The
results obtained show an excellent agreement with experimental data due to
both the exactness of the method and the accuracy of the interatomic
potential between helium atoms. However, these calculations use as
importance sampling trial wave function a Nosanow-Jastrow (NJ)
model,\cite{nosanow}  which is
not symmetric under the exchange of particles. This is not a drawback for
calculations of the energy or the spatial structure since the exchange
energy has been estimated to be negligible, about
3$\mu$K/atom,\cite{ceper1}  but the lack
of a correct quantum symmetry does not allow for estimations of the
one-body density matrix and the superfluid density. Two different models
have been proposed to overcome this limitation. The first one is inspired
in the Bloch function of band theory currently used for electrons; it is
effectively a  symmetric trial wave function but has proven to be worse
than the NJ model at the variational level.\cite{simetric}  The second one
corresponds to the shadow wave function,\cite{shadow}  which allows for a good
variational description of the solid, but not yet implemented in a DMC
calculation.

In the present work, we present a microscopic study by means of the DMC
method of bulk $^4$He immersed in a lattice of confining potentials. By
increasing progressively the strength of these external potentials we can
get relevant information on the interplay between localization and
superfluidity. Measuring the degree of localization through the Lindemann's
ratio $\gamma$ one can compare the physics of this structured phase with the solid
one. At the experimental value for $\gamma$ near the melting point
the present model shows a non-zero superfluid fraction, compatible with 
the existence of a supersolid $^4$He phase at zero temperature.

In the following section, the method used for the calculation is briefly 
described.
Sec. III comprises the results obtained, and finally, Sec. IV summarizes
the work and presents the main conclusions.

\section{Method}

The $N$-body system under study is described by the Hamiltonian
\begin{equation}
H=-\frac{\hbar^{2}}{2m}\sum_{i=1}^{N}{\bf \nabla_{i}}^{2} + 
\sum_{i<j}^{N} V(r_{ij}) +  \sum_{i,j}^{N} V_{\rm c}(|{\bf r_{i}} - {\bf
R_{j}}|) \ ,
\label{eq:hamiltonian}
\end{equation} 
with $V(r)$ the interatomic potential between He atoms, and 
\begin{equation}  
V_{\rm c}(r)= -V_0 \, \exp(-\alpha r^{2}) 
\label{eq:extpoten}
\end{equation} 
are potential wells centered in the positions ${\bf R}_{j}$, which correspond to
the sites of a solid lattice. The ground-state energy and other relevant 
properties of the system at zero temperature can be calculated exactly,
within statistical errors, using the diffusion Monte Carlo (DMC)
method.\cite{boro} 
With DMC, one is able to solve stochastically the $N$-body imaginary-time
Schr\"odinger equation by performing a random walk of the walkers
($\bf{R}\equiv\{\bf{r}_1,\ldots,\bf{r}_N\}$), which
define the wave function, according to a short-time approximation for the
Green's function.   

The simulation has been carried out by placing the external potentials
$V_{\rm c}(r)$ on the sites of an fcc lattice and with a number of
particles ($N=108$) commensurate with the lattice. At $T=0$, $^4$He crystallizes in
the hcp lattice but the differences with the fcc geometry are known to be
very  small and would not modify the conclusions of the present work. The
interatomic potential $V(r)$ is the  accurate HFD-B(HE) Aziz
model~\cite{aziz} and the
trial wave function used for importance sampling is a Jastrow factor
$\psi=\exp [\sum_{i<j}^{N} u(r_{ij})]$, with a McMillan correlation 
$u(r)=-0.5 (b/r)^5$. The parameter $b=1.3\sigma$ ($\sigma=2.556$\AA) 
is the same for all the simulations and the attractive potentials on the
sites (\ref{eq:extpoten}) are varied on its strength and range among the values
 $V_0=(500-800)$K and $\alpha=(1-2)\sigma^{-2}$. These
very sharp potentials are required to confine the atoms around the sites at
a quantitative level comparable with the one of solid $^4$He. It is worth
noticing that the localization arises due only to the branching term in the
DMC algorithm since we have chosen not to include confining terms in the
trial wave function to ensure a fully symmetric wave function. In order to
reduce any residual bias in the calculation a preliminary study on the
time-step and on the number of walkers has been carried out. In the range
of confinement simulated, the energies are well behaved and their statistical
errors are smaller than 1\%.

\section{Results}

The influence of confinement on the system has been studied at a density
$\rho=0.491\sigma^{-3}$, which corresponds to an experimental 
pressure of $\sim 34$ bar, close to the melting point of solid $^4$He.
In Fig. 1, we show a projection on the $x-y$ plane of imaginary-time
trajectories of some walkers and for a given evolution time for both the
present model and a DMC simulation using the NJ trial wave function
(DMC-NJ) at the same localization
degree. As one can see, the displacements of the atoms around the lattice
sites are in both cases rather large according to the quantum nature of
$^4$He. The difference between both simulations relies on
the exchanges between atoms at neighboring sites, that are observed in the
confining model, and which are forbidden by construction in the 
NJ description. With this configuration of the confining
potentials the probability of hopping, i.e.,  
the quotient between the number of MC steps where a hop is
produced and the total number of steps along a simulation, is $\sim 3$\%.      

\begin{figure*}
\centerline{
        \includegraphics[width=0.35\linewidth,angle=-90]{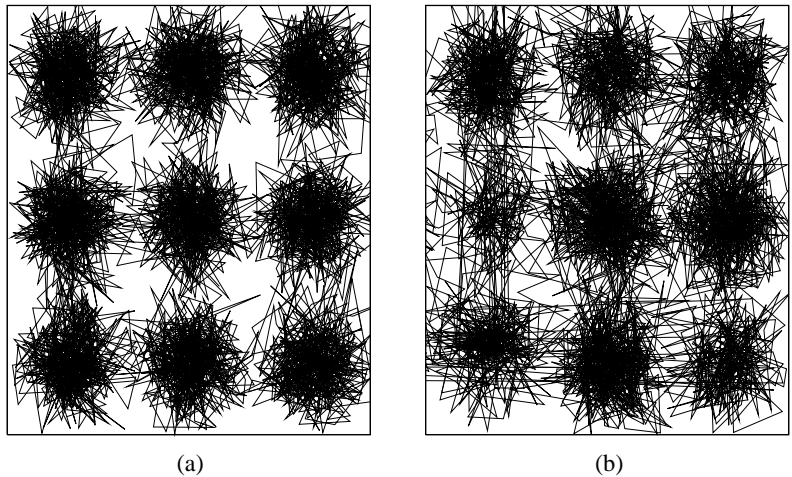}}%
        \caption{$x-y$ 
	Projection of imaginary-time trajectories along a DMC
	simulation: (a) with a N-J trial wave function; (b) with the
	present model at the same localization ($\gamma$) than case (a).  }
\end{figure*}

\begin{figure}
\centerline{
        \includegraphics[width=0.8\linewidth]{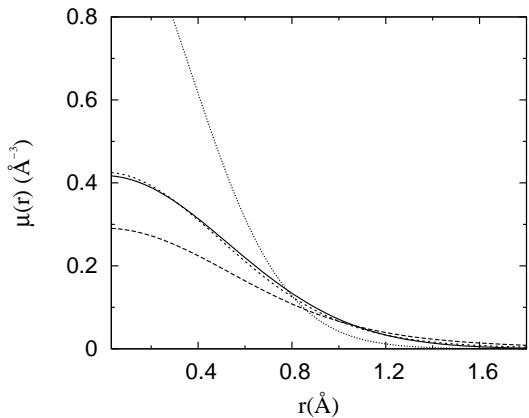}}%
        \caption{Density profile, $\mu(r)$, of the $^{4}$He atoms 
around the site positions $\lbrace {\bf R_{j}}\rbrace$. 
The solid line corresponds to the DMC-NJ
model for solid $^4$He ($\gamma = 0.257$).
Dotted, dashed, and long-dashed lines stand for the present model using
potentials $V_c(r)$ that yield  $\gamma = 0.192(1), 0.264(2)$ and 0.294(3),
respectively.}
\end{figure} 

In order to characterize the degree of localization as a function of the
range and strength of $V_c(r)$ we have calculated the density profile $\mu(r)$  
of $^4$He atoms around the lattice sites $\lbrace {\bf
R_{j}}\rbrace$. From them, one can estimate the mean squared displacement 
\begin{equation}
  \langle u^{2} \rangle = 4 \pi \int_{0}^{\infty} \mu(r) r^{4} dr \ ,
\label{eq:msdis}
\end{equation}
and then the Lindemann's ratio 
$\gamma=\sqrt{\langle u^{2} \rangle}/a$, $a$ being the nearest-neighbor distance
in the lattice. In Fig. 2, results for $\mu(r)$ for three confining models
are plotted in comparison with the DMC-NJ result of solid $^4$He.
The values for $\gamma$ of the three cases are $0.192(1)$, $0.264(2)$, and  
$0.294(3)$, to be compared with the DMC-NJ value $\gamma = 0.257$  which agrees
with the experimental value $\gamma^{\rm expt} = 0.26$.\cite{burns}  All the lines in
Fig. 2 are fits to the DMC results using a sum of two Gaussians, a simple
model which has proved to reproduce accurately the computed data. The
figure shows that the present model is flexible enough to simulate from a
strongly confined system, with $\gamma=0.192(1)$, until a more delocalized
one, with $\gamma=0.294(3)$, crossing the value of the Lindemann's ratio of
solid $^4$He. As one can see, the curve of the model for $\gamma= 0.264(2)$
in nearly indistinguishable from the DMC-NJ result.

\begin{figure}[t]
\centerline{
        \includegraphics[width=0.8\linewidth]{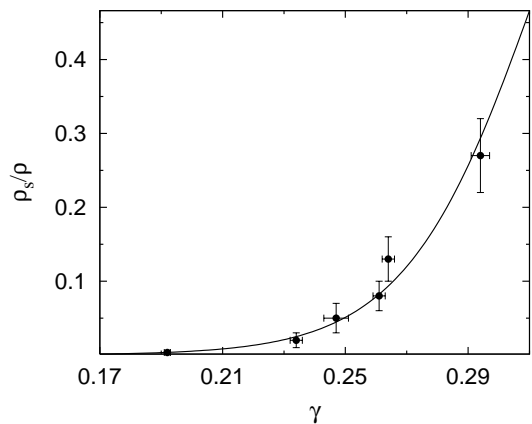}}%
        \caption{The superfluid fraction as a function of the Lindemann's
	ratio $\gamma$ in 
	the localized model. The line is a fit (\protect\ref{fitrho}) to 
	the computed data.
	}
\end{figure}

The main objective of the present work is the estimation of the superfluid
density as a function of the degree of localization of particles around the
lattice sites. 
The superfluid density of a bosonic system may be computed
with DMC by extending the winding-number technique, originally developed
for PIMC calculations, to zero temperature.\cite{zhang}  Essentially,  this is obtained
by calculating the diffusion constant of the center of mass of the
particles in the simulation box for an infinite imaginary time 
\begin{equation} 
\frac{\rho_{\rm s}}{\rho}=\lim_{\tau \to \infty}
\frac{1}{6 N \tau} \left(\frac{D_{\rm s}\left(\tau\right)}{D_{0}}\right) 
\ ,
\label{eq:superdens}
\end{equation} 
where $D_{\rm s}(\tau)=\langle({\bf R_{\rm CM}}(\tau)-{\bf
R_{\rm CM}}(0))^{2}\rangle$, and $D_{0}=\hbar^{2}/2m$. Results obtained
for $\rho_{\rm s}$  corresponding to several configurations of the
lattice potentials $V_{\rm c}$ are plotted in Fig. 3. In this figure,  
$\rho_{\rm s}/\rho$ is shown as a function of the Lindemann's ratio of the 
same configuration and for values  $0.18<\gamma <0.30$. The results for 
$\rho_{\rm s}(\gamma)$ are well parameterized by the function
\begin{equation}
\rho_{\rm s}/\rho (\gamma) = \left[ 1 + \exp \left( -\frac{\gamma -
\gamma_0}{\omega} \right) \right]^{-1} \ ,
\label{fitrho}
\end{equation}
shown as a line in Fig. 3. The parameters in Eq. (\ref{fitrho}) are
$\gamma_0=0.313(4)$ and $\omega=0.021(1)$. From Eq. (\ref{fitrho}) one can
estimate the prediction of our model for the range of more interesting values
for $\gamma$. In particular, considering $\gamma^{\rm expt}=0.26$,  
the superfluid fraction is $ \rho_{\rm s}/\rho = 0.079(16)$.  It is worth
noticing that the present results for $\rho_{\rm s}/\rho (\gamma)$ are
qualitatively similar to the ones obtained by Saslow~\cite{saslow} assuming an
uncorrelated model where all particles develop a common phase function.
However, his most recent calculations predict smaller values for the
superfluid fraction than ours, and in particular, for solid $^4$He a value 
$ \rho_{\rm s}/\rho=0.022$.

The connection between the localized model here presented and the NJ
description of solid $^4$He has been made so far by comparing the Lindemann's
ratio of both calculations. To make this comparison more complete, 
we have explored if there are other structure properties which can also 
be comparable for similar values of $\gamma$. To this end, we have computed
the two-body radial distribution function $g(r)$, which is proportional to
the probability of finding two particles a distance $r$ apart, and its
Fourier transform, the static structure factor $S(k)$. 

In Fig. 4, results of $g(r)$ for two different configurations of the
confining potentials $V_c(r)$ are shown in comparison with the DMC-NJ result at
the same density. These functions, as also the density profiles $\mu(r)$
shown in Fig. 2, have been calculated using a pure estimation technique to
avoid the residual bias coming from the trial wave function that the direct
simulation output (mixed estimation) presents. As one can observe in the
figure, when the Lindemann's ratio of the confining model is very close to
the one of the NJ model both curves coincide. In the same figure, a result
corresponding to a smaller $\gamma$ value shows clearly a more pronounced
peak reflecting its higher localization around the lattice sites.

\begin{figure}[t]
\centerline{
        \includegraphics[width=0.8\linewidth]{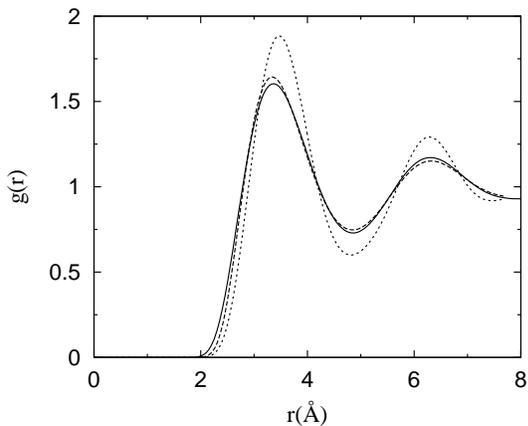}}%
        \caption{Two-body radial distribution function of the localized
	model corresponding to superfluid fractions $ \rho_{\rm s}/\rho
	=0.08$ (long-dashed line) and 0.035 ( short-dashed line). The solid
	line stands for the DMC-NJ result at the same density.
		}
\end{figure}

\begin{figure}[t]
\centerline{
        \includegraphics[width=0.8\linewidth]{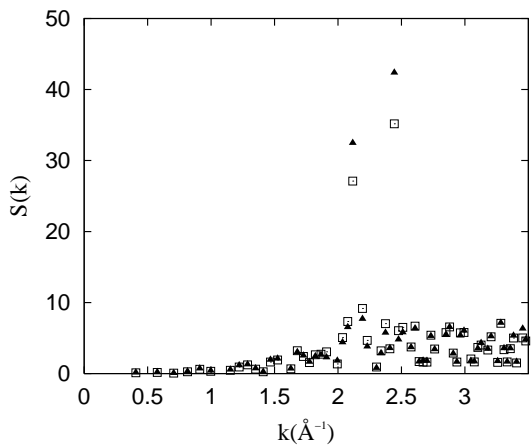}}%
        \caption{Static structure factor of the localized model
	corresponding to $ \rho_{\rm s}/\rho =0.11$ and $\gamma=0.264$
	(squares). The DMC-NJ result is also shown for comparison
	(triangles).
		}
\end{figure}

The solid-like character of a given configuration is much more clear by
looking at the static structure function $S(k)=1/N \langle \rho_{-\bf k}
\rho_{\bf{k}} \rangle$, with $\rho_{\bf k} = \sum_{i=1}^{N} e^{i {\bf
k}\cdot {\bf r}_i}$, than $g(r)$. In Fig. 5, results of $S(k)$ for the
present model are compared with the DMC-NJ result for solid $^4$He. The
localized model result shown in the figure corresponds to a Lindemann's
ratio $\gamma=0.264(2)$, close to the DMC-NJ value for $\gamma$, 
and a superfluid fraction  $ \rho_{\rm s}/\rho =
0.11(1)$. Both results show the presence of the two main first peaks at the same
$k$ values, characteristic of the underlying fcc lattice. The strength of
the peaks of the localized model is slightly lower than the one of the DMC-NJ
calculation since the values for $\gamma$ in both models are not exactly
the same, but it is clear that the present model is able to reproduce the
expected trend of a solid phase in $S(k)$.

A slight trend of increasing $\rho_{s}/\rho$ with pressure is evidenced on
Kim and Chan's experiments,\cite{moses1}  contrarily to what it occurs in superfluid
$^{4}$He; the authors conjecture that this may be due to the presence of
grain boundaries formed in the growth of the crystal  and which affect the
coherence in the superflow.
Regarding this issue, we have computed $\rho_{s}/\rho$  
at higher density, $\rho_{\rm u}=0.529\sigma^{-3}$, and external constraint
yielding $\gamma=0.264(2)$ and $\rho_{s}/\rho=0.13(3)$ at density
$\rho = 0.491\sigma^{-3}$.  The pressure of bulk solid
$^{4}$He at density $\rho_{\rm u}$ ($\sim 64$ bars)  is still on the range of
supersolid experiments. We find $\rho_{s}/\rho=0.10(2)$ and
$\gamma=0.239(5)$  ($\gamma=0.246(3)$ for the DMC-NJ simulation at $\rho_{\rm
u}$). Therefore, within the statistical uncertainties of the model 
we do not observe pressure dependence of $\rho_{s}/\rho$ in the range of
pressures analyzed.

\section{Conclusions}

We have studied the competing effects of localization and
superfluidity of bulk $^4$He immersed in a lattice of external confining
potentials. The simulations have been carried out  at zero temperature 
using the DMC method. From
the curve $\rho_{\rm s}/\rho(\gamma)$, and using the experimental value of
$\gamma$ of solid $^4$He we obtain an estimation of the superfluid fraction
$ \rho_{\rm s}/\rho = 0.079(16)$, a value significantly larger than the
experimental measure ($ \rho_{\rm s}^{\rm expt}/\rho = 0.01-0.02$). 
On top of the possible drawbacks of the localized  model for describing the $^4$He
solid phase, there are two effects
that could help to explain the difference between the experimental measure
and the present prediction. They are, first, the temperature (the
simulation is at zero temperature), and second, the possible influence of
impurities in the experiments. The latter can be specially relevant since,
as commented by Kim and Chan,\cite{moses1,moses2} a crucial point to observe
superfluidity is the as much as possible elimination of impurities in the
sample. The
influence on the superfluid signal of $^3$He impurities and/or
incommensurability 
can be also analyzed in the present framework. Work is in progress along
this direction.

\acknowledgments
We acknowledge partial financial support from DGI (Spain) Grant No.
FIS2005-04181
and Generalitat de Catalunya Grant No. 2005SGR-00779.

\end{document}